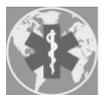

# Time granularity impact on propagation of disruptions in a system-of-systems simulation of infrastructure and business networks

**Mateusz Iwo Dubaniowski [1\*], Hans Rudolf Heinimann [1]**

[1] ETH Zurich, Future Resilient Systems, Singapore-ETH Centre; iwo.dubaniowski@frs.ethz.ch
\* Correspondence: iwo.dubaniowski@frs.ethz.ch

**Abstract:** System-of-systems (SoS) approach is often used for simulating disruptions to business and infrastructure system networks allowing for integration of several models into one simulation. However, the integration is frequently challenging as each system is designed individually with different characteristics, such as time granularity. Understanding the impact of time granularity on propagation of disruptions between businesses and infrastructure systems and finding the appropriate granularity for the SoS simulation remain as major challenges. To tackle these, we explore how time granularity, recovery time, and disruption size affect the propagation of disruptions between constituent systems of an SoS simulation. To address this issue, we developed a High Level Architecture (HLA) simulation of 3 networks and performed a series of simulation experiments. Our results revealed that time granularity and especially recovery time have huge impact on propagation of disruptions. Consequently, we developed a model for selecting an appropriate time granularity for an SoS simulation based on expected recovery time. Our simulation experiments show that time granularity should be less than 1.13 of expected recovery time. We identified some areas for future research centered around extending the experimental factors space.

## 1. Introduction

Development of new technologies results in infrastructure systems becoming more interdependent thus introducing additional complexities. These systems require and produce inputs and outputs not only for internal use by the systems themselves, but also for other infrastructure systems and businesses. Often, those businesses also provide infrastructure resources that are then delivered over a systems network. Along with these heightened interdependencies, systems disruptions are increasing in both magnitude and frequency. This is especially visible within the context of urban settings, where various interdependent systems are vital to the survival and normal operation of a society [1]. As a result, the design and development of infrastructure systems must be done in a way that ensures they are resilient and can sustain a large variety of disruptions. While a major concern of designers is the proper response to disruptions, planners and policymakers must recognize how disruptions emerging in one system can affect other systems, and how disruptions propagate from one system to another.

Currently, however, there is insufficient understanding about how such propagation of disruptions between systems occurs [2-6]. To understand this, simulations and models of infrastructure systems are often run to predict how systems behave under a disruption [6]. Similar issues appear in many other fields such as evolutionary dynamics of social economy [7-9]. Although the effect of a propagated disruption on a simulation is affected by many factors, the extent to which modeled environments are influenced has not been adequately studied [3] [10-12]. These streams of research focus on supply chain propagation of disruptions and their adequate modeling, however, they seldom incorporate infrastructure systems into these models. It is interesting to see how infrastructure systems simulations are affected by factors that influence disruption propagation. Moreover, these authors focus primarily on outlining various methods of representing disruption propagation, however, they do not consider factors, such as time granularity, that might affect disruption propagation. For example, time granularity of a simulation is expected to affect the propagation of disruptions. While time granularity of a simulation is expected to be a crucial factor, no comprehensive research has been conducted on how it might affect a propagation of disruptions between infrastructure systems and businesses within a system-of-systems (SoS) framework. Therefore, the practical value of such simulations is diminished due to their limited correspondence with real-world scenarios. Consequently, we focus on understanding the impact of time granularity on propagation of disruptions, which is vital if we are to determine how disruptions to infrastructure systems and businesses affect societies.

Although models have been designed to examine interdependencies among infrastructure systems [13-17], they have not considered the issue of time granularity and how it influences the propagation of disruptions between systems under an SoS simulation. Current investigations include separate analyses of individual systems, e.g., traffic simulation [18], water supplies [19], or power grids [20]. However, only single systems have been involved there, thus constraining those analyses that might address propagation of disruptions to several infrastructure systems. Similarly, propagation issues between systems have been studied in signal processing literature [21-24]. However, these do not look at propagation of disruptions between infrastructure systems, especially in the context of SoS simulations. Another stream of research, utilizing SoS simulations of infrastructures [25-27], does not focus on modeling disruptions or ensure the accurate capture of their propagation. In contrast, Dubaniowski and Heinimann [6] have examined the impact of time granularity on infrastructure systems. However, their study has not considered businesses or the impacts of disruption size and recovery time on the propagation of disruptions. Our study remedied this gap by including businesses in the simulation and considering the impacts of more factors such as disruption size and recovery time. Furthermore, an SoS model of infrastructure systems within an urban ecosystem, where disruptions are introduced [27-28], has limited applicability because it does not account for the propagation of those disruptions, and does not provide for many variations based on time granularity of the simulation and different types of disruptions. In this study, we tackle the challenge of understanding propagation of disruptions and its dependence on different time granularities as well as other factors such as disruption size and recovery time.

The objectives of our study were to develop a distributed system-of-systems model of infrastructure systems and businesses to: (1) study the effects of different disruption characteristics on propagation of disruptions between constituent systems; (2) investigate how time granularity of distributed model can affect propagation of disruptions in the model; and (3) develop a framework model for selecting the most appropriate time granularity of an SoS distributed model based on expected, estimated disruption parameters. In particular, our goal was to investigate how time granularity of a simulation, as well as the recovery time and size of a disruption to a theoretical constituent network – water supply – might propagate and affect the outcome for businesses that are networked within the simulation. In this study, we also present a general framework for performing such analysis on any SoS simulation of several constituent network systems.

This study aims to improve the accuracy of SoS simulations of infrastructure systems and businesses, and so the correspondence of those simulations with the real-world. Particularly, this study addresses the issues arising from combining various infrastructure system models operating at different time granularities. Inclusion of such systems in an SoS framework poses many challenges and often presents inadequate results due to disparity of time granularities between constituent SoS systems. While the choice of the overarching SoS time granularity is vital to representing disruptions propagation in the SoS model adequately, this is not addressed adequately in the current research streams as shown in the review above. Currently, disruptions represented in an SoS simulation often do not propagate properly between SoS components due to inadequate overarching time granularity of the SoS simulation. This leads to repeated simulations, which waste simulation budgets, or results in inadequate outcomes. Results of such simulations do not bring as much value because of having significant disparities with the real-world results and not representing the actual disruption events accurately. Obtaining the suitable time granularity for such simulations would bring results of these SoS simulations closer to the real-world values, what this study attempts to achieve. Consequently, as a result of this study, practitioners, as well as decision-makers, of risk and disaster management will benefit from better simulations and decision support tools, which in turn would contribute to better responses to disruptions. In particular, the gap addressed in this study is the accurate selection of the time granularity parameter in an SoS simulation. Closing this gap will lead to more accurate simulations, and thus better decisions by the users of these simulations such as infrastructure planners, policy makers, or risk managers.

Our expectation in the analysis part of this study was that time granularity would have the most significant impact on the propagation of disruptions, because coarse time granularity can completely bar disruptions of very short periods from cascading to other systems in the simulation. For this study, we did not consider a variety of networks or how their topology might affect the propagation of disruptions. The cause of the initial disruption was also outside the scope of this study.

The main contributions of our study were: 1) the development of a framework to select the appropriate time granularity for an SoS simulation of disruption propagation; 2) application of this framework to infrastructure and business networks resulting in a recommendation regarding the choice of the optimal time granularity, which according to our findings should not exceed 1.13 of the expected recovery time of the disrupted system. The findings can be further extended to other fields such as finance, biology, evolutionary dynamics of social economy.

This study takes up the challenge of investigating the impact of time granularity on propagation of disruptions in SoS simulation of infrastructures and businesses. The rest of this paper is organized as follows. Section 2 describes in detail the model specification, particularly the conceptual framework that we use to model infrastructures and businesses. In Section 3, we outline the set up that we use in the application of the framework for the simulation experiment, and the implementation of the systems. In Section 4, we present the results of the simulation experiments with regards to different metrics, as well as model and recommendation as to time granularity in SoS simulations of infrastructures and businesses. Finally, in Section 5, we present the conclusions of this study, particularly the key findings and implications of our study and an overview of future work that could be conducted on this subject.

## 2. Conceptual framework – system-of-systems of infrastructure systems

Frameworks and methodologies have been established to model individual infrastructure systems and businesses, e.g., power [20] or water supplies [19], transportation [29], emergency services [30], or financial systems [31]. Those models correspond only to individual infrastructure systems, and are independent and autonomous in the way they represent each separate system. However, in reality, these systems are interconnected due to various interdependencies among infrastructure systems and the businesses to which they deliver. For example, water supply systems are heavily dependent on a power supply to operate their pumps, and emergency services rely upon both power and water to run hospitals that treat sick people. Those people must also be moved to and from hospital over transportation networks. Similarly, a business such as a restaurant is dependent on access to power to operate its machinery and kitchen equipment, and on water supply to cook and serve meals and clean the equipment.

The interdependencies between infrastructure systems and businesses can be modelled in an overarching framework. The SoS approach models individual systems as being autonomous in their internal operations, but at the same time connected with and affected by other systems [25][27][32]. Therefore, this approach considers both inter- and intra-system interdependencies. In such a framework, infrastructure systems and businesses are standalone models, while the interdependencies between them are simulated as lifeline connections (Figure 1). Those lifelines provide vital infrastructures and businesses with access to network systems, thereby mimicking their interdependencies.

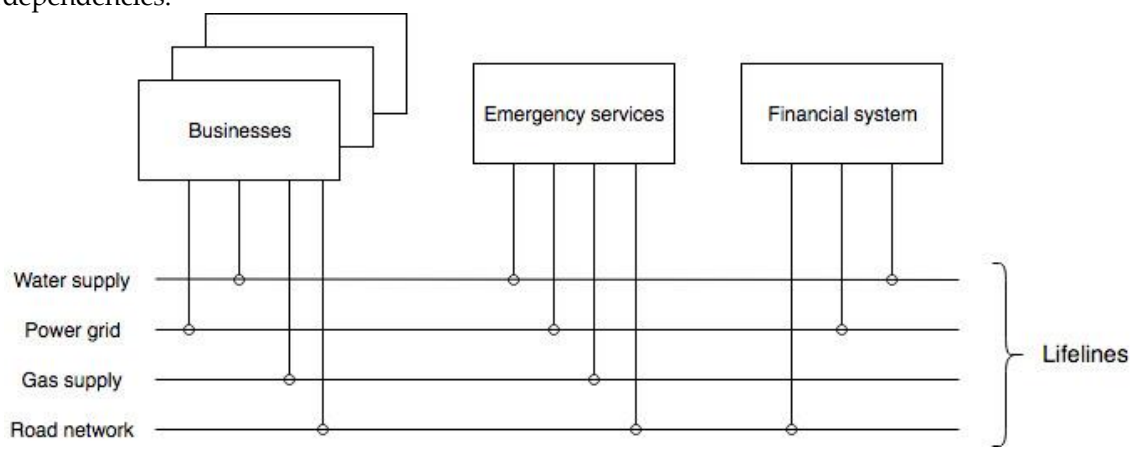

**Figure 1.** Conceptual model of infrastructure system-of-systems, where complexity is two-fold, i.e., within and between specific systems.

The conceptual model shown on **Figure 1** represents good duality under real-life conditions, with individual systems also linked through roads, power lines, water and gas pipes, and similar infrastructure networks. Services are delivered to provide access to and distribution of actual infrastructure systems and the resources produced by businesses. On the Figure, the interdependencies are exhibited through lifeline infrastructure connections between various constituent systems. A system depends on lifelines that it is connected to for performing this system's critical functions.

In this study, time granularity is defined as the frequency of performing synchronization between constituent systems – federates – of an SoS simulation, expressed as a number of timesteps between two consecutive synchronizations of all federates. The concept of time granularity is of great importance in a simulated SoS setting [6] because the impact and propagation of disruptions between constituent systems can vary significantly depending on the time granularity of that simulation. Therefore, we developed SoS simulation experiments of infrastructure systems and business networks combined with a disruption generator. These experiments allowed us to understand how time granularity affects propagation of disruptions in the SoS.

To perform the analysis of how time granularity and other factors affect propagation of disruptions between systems within a context of an SoS simulation, and consequently to choose an appropriate time granularity, we have developed the following general framework (**Figure 2**):

1. Select number and types of individual systems to be modeled, define their performance metrics, and experimental factors to be tested.
2. Identify interdependencies between selected systems and define these interdependencies.
3. Implement individual systems' networks into an SoS HLA simulation.
4. Implement disruption introduction mechanism, include this mechanism in the SoS HLA simulation.
5. Fix the time available for the simulation experiment, hence determine the number of experiments that will be executed and the experiment layout e.g. full-factorial layout or less dense layout.
6. Run the simulation experiment and record the data.
7. Perform ANCOVA analysis on the data to find factors affecting disruption propagation the most, and to determine the relationship of these factors to time granularity.
8. Draw conclusions as to how time granularity of the SoS simulation should be selected for this type of simulation with regards to simulation parameters and desired accuracy.

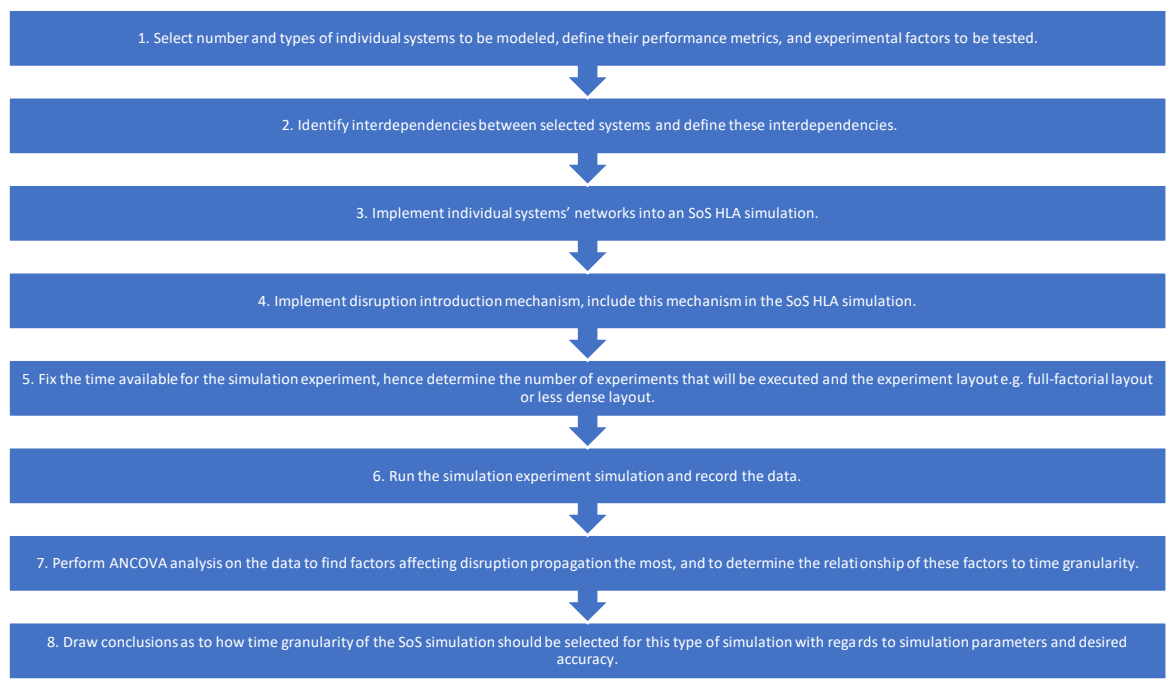

**Figure 2:** Framework for the analysis of time granularity effects on propagation of disruption in an SoS simulation.

The above steps show a general procedure to follow in order to analyze how propagation of disruptions between constituent systems of an SoS simulation is affected by different factors. This procedure allows to gather information on how to select an appropriate time granularity for the simulation. Furthermore, interactions of factors with time granularity can be tested too to understand their impact on propagation of disruptions. As a result, valuable information on how to select crucial parameters for the simulation can be obtained. This is of tremendous value when designing large scale SoS simulations. Small-scale prototype simulations such as presented in this study can be used to establish the range of values that time granularity should take for large-scale simulations of similar networks, and what factors should affect the selection of time granularity.

In this section, we described the specification of the model used in the simulation, we outlined the framework to follow to derive an optimal time granularity for SoS simulation. In the next section, we focus on the experiment that we performed by applying the above framework.

## 3. Methodology – experimental setup

In the previous section, we described the model in detail and we described the conceptual framework that we used to derive the experimental simulation described in this section. In this section, we focus on describing the experimental setup i.e. layout, metrics, and implementation in detail.

### 3.1. Experimental layout

We designed an experimental model system to examine the change in disruption patterns for constituent systems as a function of time granularity (**Figure 3**). When combined with a disruption generator, we could introduce system disruptions in accordance with prescribed patterns. Our observer module was then used to visualize results in real time as the simulation progressed so that we could determine how the impact of the disruption was propagated within the system over time.

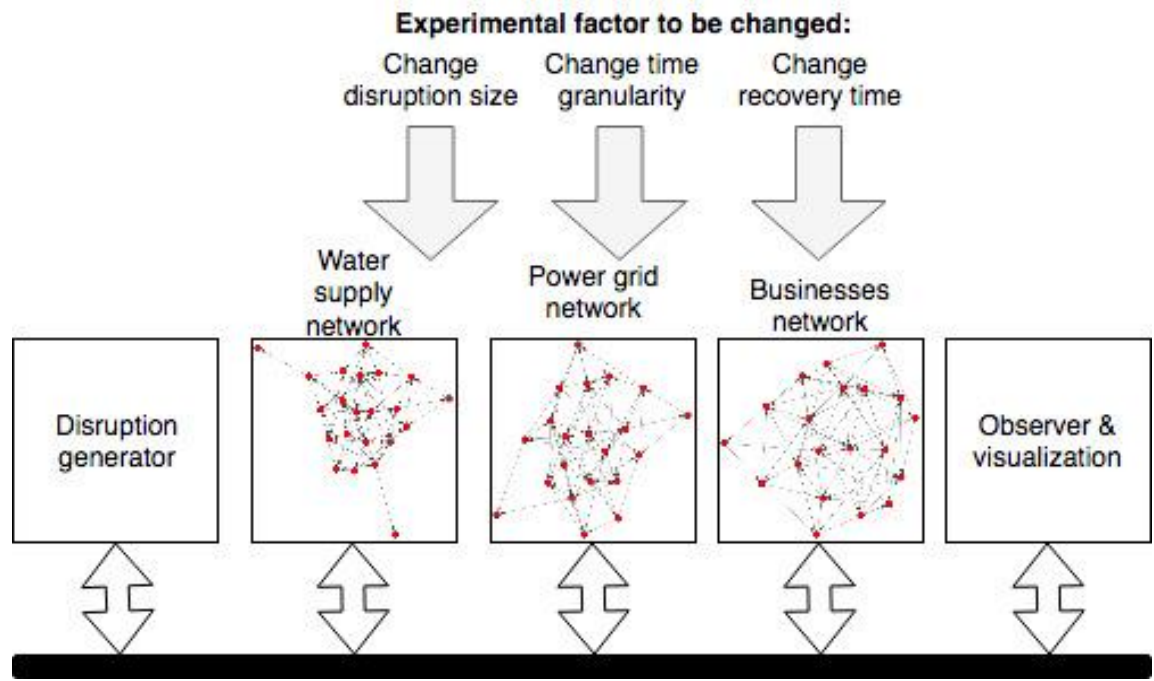

**Figure 3.** Experimental model system setup for 3 networks and disruption generator.

The three networks were abstractly generated based on their unique characteristics, as well as on their interdependencies and corresponding connections to each other. This process is described in more detail in Section 3.1.2 below. Synchronization between networks occurred at a predetermined frequency, i.e., federates performed simulations internally in timesteps without any wait times between subsequent timesteps. However, at the synchronization points, the federates had to halt and wait for all other federates to catch up so that their data could be exchanged.

The three networks included water supply network, power grid network and businesses network. All these networks have unique characteristics. Water supply and power grid networks tend to have branch network topology. Their reaction time is different with water network reacting to disruption within minutes and power grid within milliseconds, On the other hand, a businesses network has a branch topology too, however, with many links that connect to others without going through the common root i.e. with hints of mesh topology. This is complemented by a lot of interdependencies to both water supply and electric grid network. The propagation of disruption within the network happens in matter of minutes to hours, and this is reflected in the model by appropriate delays. Hence, the propagation of disruptions from other networks if these appeared could be captured in the businesses network.

After disruptions were introduced into the water supply network, their propagation through the SoS was assessed according to the impact they had on the business network. As shown in **Figure 3**, three experimental factors were varied: time granularity, recovery time after the disruption, and the disruption size. For our purposes, the disruption generator followed Poisson processes, which adequately represent real-world disruption occurrences [33-34]. The two parameters used in defining our original disruption included (1) actual disruption size (DS), i.e., the number of affected nodes in water supply network; and (2) actual recovery time (RT), which indicated how long the disruption remained effective in the water supply network. For the actual disruption size, we randomly generated a particular pattern of affected nodes for each disruption size parameter level based on random number generator. We have used this generated pattern throughout all experiment configurations. This allowed us to shape the disruption curve and compare the size of a disruption and its propagation and recovery pattern depending on different simulation parameters and experimental factors' values.

The electric grid network was introduced into the simulation to show how independent simulation that itself does not introduce disruption to the SoS can affect the propagation of disruptions between two other networks. The electric grid showed how a system that is not disrupted initially can become affected by a disruption originating in another system, and subsequently also additionally impact the observed businesses network. The main objective of the study was to understand time granularity effects on disruption propagation characteristics. Hence, electric grid was not measured itself because it would not bring significant additional value to this study, however, the process could be similarly repeated for the electric grid network as the target network of our observations.

We also determined how time granularity might affect simulation speed. Because synchronization of the system required time, increasing the frequency of synchronizations, i.e., decreasing time granularity, would decrease the speed of simulation. Our goal was to understand the trade-off between accuracy and speed so that we could choose the most appropriate time granularity parameter for the SoS simulation and, hence, estimate the runtime of the simulation under such granularity.

To sum up, the experimental process was as follows:
1. Develop 3 networks with interdependencies between those: water supply, power supply, businesses; described in Section 3.1.2.
2. Introduce a disruption to water supply simulation. Disruption and experimental configurations are described in more detail in Section 3.1.1.
3. Register performance metrics (described in Section 3.2) of the businesses network simulation.
4. Repeat the above for each experimental configuration described in Section 3.1.1 (**Table 1**).

The implementation of the above is described in Section 3.3. While conducting the above experiment, we were interested in recording propagation of disruptions from one infrastructure system – water supply network – to businesses network. Hence, we focused on recording metrics related to the propagated disruption i.e. propagated disruption size and propagated recovery time. These were analyzed in relation to time granularity in order to see the impact that time granularity of the SoS had on propagation of disruptions between constituent networks. We developed a model for deriving an adequate time granularity for the SoS simulation such that the propagation of disruptions would register. Finally, we looked at the trade-off between simulation time and time granularity. This emphasized the importance of selecting an appropriate time granularity under limited simulation budgets.

In this study, we present an application of the above procedure to an SoS consisting of 3 systems and we test impact of 3 experimental factors on propagation of disruptions between constituent systems of an SoS simulation. The rest of this study describes this scenario in more detail and outlines the application of the above procedure to the scenario, including the analysis of results with inferences on how to select an appropriate time granularity for an SoS simulation. However, the general framework in principle can be applied to any combination of infrastructure systems and other network systems in order to determine the impact of various factors on propagation of disruptions between constituent systems of an SoS simulation.

3.1.1. Factorial layout

We applied a full-factorial experimental layout to study the impact of time granularity (TG), actual recovery time (RT), and disruption size (DS) on simulation results (**Table 1**). Those three factors were assigned values based on Latin Hypercube Sampling (LHS) [35]. Because the overall water supply network size was 22 nodes, sampling for disruption size was performed in the space between 7 and 21 nodes disrupted. Time granularity and recovery time were both assessed on the space of between 1 and 30 to provide us with a good overview of real-life simulations. The full factorial experimental layout consisted of 125 parameter configurations. This allowed us to identify solid conclusions by which we could determine the impact of individual factors on the accuracy and outcome of the simulation.

**Table 1.** 5 x 5 x 5 hypercube full-factorial layout achieved via LHS. Experimental factors included time granularity (TG), recovery time (RT), and disruption size (DS).

| | | Disruption size (DS) | | | | | | | | | | | | | | | | | | | | | | | |
|---|---|---|---|---|---|---|---|---|---|---|---|---|---|---|---|---|---|---|---|---|---|---|---|---|---|
| | | 8 | | | | | 12 | | | | | 14 | | | | | 18 | | | | | 21 | | | | |
| | | **Time granularity (TG)** | | | | | **Time granularity (TG)** | | | | | **Time granularity (TG)** | | | | | **Time granularity (TG)** | | | | | **Time granularity (TG)** | | | | |
| Recovery time (RT) | 2 | 2 | 12 | 14 | 21 | 27 | 2 | 12 | 14 | 21 | 27 | 2 | 12 | 14 | 21 | 27 | 2 | 12 | 14 | 21 | 27 | 2 | 12 | 14 | 21 | 27 |
| | 9 | 2 | 12 | 14 | 21 | 27 | 2 | 12 | 14 | 21 | 27 | 2 | 12 | 14 | 21 | 27 | 2 | 12 | 14 | 21 | 27 | 2 | 12 | 14 | 21 | 27 |
| | 13 | 2 | 12 | 14 | 21 | 27 | 2 | 12 | 14 | 21 | 27 | 2 | 12 | 14 | 21 | 27 | 2 | 12 | 14 | 21 | 27 | 2 | 12 | 14 | 21 | 27 |
| | 17 | 2 | 12 | 14 | 21 | 27 | 2 | 12 | 14 | 21 | 27 | 2 | 12 | 14 | 21 | 27 | 2 | 12 | 14 | 21 | 27 | 2 | 12 | 14 | 21 | 27 |
| | 22 | 2 | 12 | 14 | 21 | 27 | 2 | 12 | 14 | 21 | 27 | 2 | 12 | 14 | 21 | 27 | 2 | 12 | 14 | 21 | 27 | 2 | 12 | 14 | 21 | 27 |

3.1.2. Specification of topologies for infrastructure systems

Networks used in this simulation experiment were abstract, and were randomly generated via the Erdős-Renyi model [36-37]. The use of real-world networks would improve the significance of the study and is intended to be included in the future. However, at the moment we lack access to data representing accurate networks in the same area with the relevant time granularity. Furthermore, the development of a real-world case study for this project would far exceed the time and resources available. The primary purpose of our manuscript is to highlight the issue of time granularity in developing SoS dynamic simulations, and to present a basis and framework for the selection of an appropriate time granularity in these simulations. In the future, we do hope to expand on the issues mentioned in our manuscript and eventually to deliver a real case study. The networks included in our study represented water supply, power grid, and businesses. They were represented as graphs in which direction signified the way in which interactions would occur between subsequent nodes, such as those associated with normal transportation of a resource over various links. Disruptions were introduced at nodes, which were then removed from the network and their connection links abandoned. Recovery was simulated by returning those nodes to the network and re-establishing their connections with subsequent and preceding nodes.

In each network, nodes corresponded to units that performed operations and interacted with other nodes in the same network as well as with corresponding nodes in other federates. Edges corresponded to transfer links between operational units within each network. Although the mechanics of each network were similar, they were also distinct and abstract. Each node had its own intrinsic, internal performance, but also took inputs from the incoming edges of its network and from its corresponding nodes in the other two networks. These internal and external performances were then combined and transformed to determine the total performance of that particular node. Performance was then propagated to the following nodes through the outgoing edges. Similarly, performance was propagated through inter-network connections and a synchronization mechanism to the corresponding nodes of the other two networks. Interaction between different federates could take place only through synchronization. In doing so, we designed a working process for each constituent federate to simulate an individual infrastructure system or business network within the SoS simulation environment. Each network had similar but slightly different mechanics for calculating the performance of its own nodes.

A section of network topologies is shown in **Figure 4**. The Figure shows selection of nodes and their connections for the networks used in the simulation experiment. Each network consisted of a certain number of nodes with a certain number of edges between them. A selection of these nodes and edges is shown on **Figure 4**. The nodes also had corresponding interdependent nodes in other networks of the SoS simulation with which they communicated at a given frequency by exchanging information at synchronization points. These interconnections between networks are represented on **Figure 4** with the dashed lines. We set the following specifications for the networks used in the experiment: 22 nodes and 77 edges for the water supply network, 21 nodes/77 edges for the power grid network, and 20 nodes/75 edges for the business network.

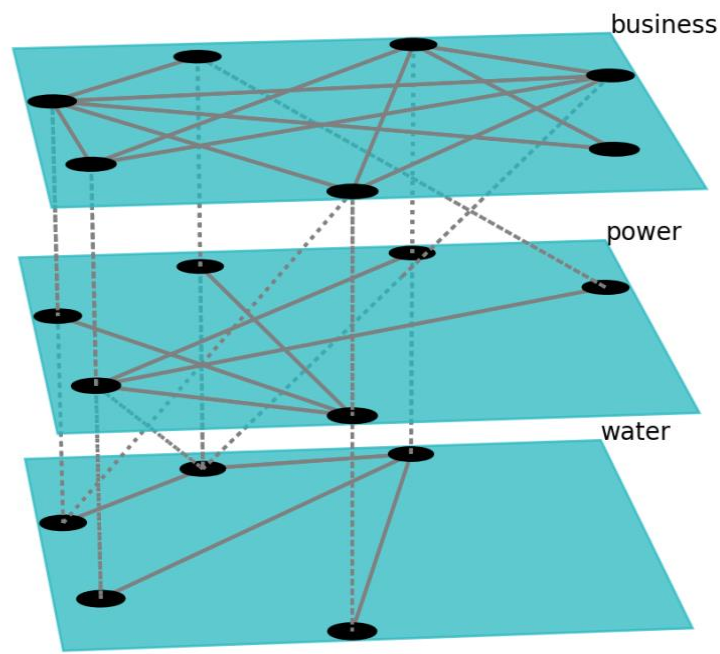

**Figure 4.** Section of topologies of 3 simulation experiment networks.

*3.2. Performance metrics*

We collected data that described how a disruption to the water supply might influence performance of the business network. The collected data included simulated, propagated recovery time (SPRT) and simulated, propagated disruption size (SPDS) of the business network (Table 2). These metrics were chosen as the most representative due to showing the most important features of a disruption. Our aim was to derive an independent framework for choosing time granularity for an SoS simulation of infrastructure systems and business networks that would result in the most accurate simulation while simultaneously preserving the efficiency of that simulation. We used a Measure of Performance (MoP) to compare the results from experimental configurations, based on sum of all individual performances of nodes in a network. We expressed MoP as a percentage of its default, optimal conditions performance before the disruption strikes, which was set at 100%. MoP can be understood as the overall performance of the system and can be compared across different instances of time in order to see how performance of the system is affected by experimental factors. MoP for each network is obtained by summing all individual node's performances for that network. For example, MoP for a water network could correspond to the sum of all the water delivered to water consumers in the region e.g. expressed in $m^3$. Similarly, in case of a power system it can correspond to the total electricity transported in the network e.g. expressed in Wh. For businesses, the MoP can represent the total economic activity in the area such as GDP or total value of goods and services produced in that area e.g. expressed in $. Such metrics succinctly show any disturbances in the respective networks by decreased production and delivery of resources, which would be required under normal operating conditions.

**Table 2**. Performance metrics and variables

| Performance metric | Description |
|---|---|
| Simulated propagated recovery time (SPRT) ($t_{rec}$) [timesteps] | The length of time needed for a system to recover to 99% of its original performance after a disruption was propagated and, subsequently, retracted. Expressed in timesteps. |
| Simulated propagated disruption size (SPDS) ($d_{max}$) [%] | The difference between 100% and the lowest point along the MoP curve of the target system after a disruption was propagated to that system. |
| Actual, initial recovery time in the original system (RT) [timesteps] | The length of the disruption in the system in which the disruption originated i.e. in the water supply network. Expressed in timesteps. |
| Actual, initial disruption size in the original system (DS) [%] | The difference between 100% and the lowest point along the MoP curve of the initial system after a disruption was introduced to that system |
| Measure of performance (MoP) [%] | Measure of satisfaction of the demand for resources in the system. Expressed as percentage of the normal operating conditions, when all of the demand is satisfied. |
| Time granularity (TG) [timesteps] | Frequency of synchronization of the systems i.e. frequency of exchange of interdependent data between constituent systems in the SoS simulation. |

Disruptions and recovery are modelled in a similar fashion. Disruption is modelled by decreasing the performance of an individual node or a set of nodes sharply in one timestep to null performance in effect by removing the node from the network, and subsequently observing how the target network reacts to this. Similarly, recovery is modelled by reverting the disrupted nodes in one sharp timestep back to their original levels of performance from before the disruption had occurred, and subsequently observing how the target network rebounds and recovers from the disruption. Individual network mechanisms and within-network as well as between-network propagations are responsible for the recovery process.

As depicted in **Figure 5**, the simulated propagated recovery time ($t_{rec}$) (SPRT) enabled us to calculate the length of time needed for a system to recover after a disruption was propagated and, subsequently, retracted. Recovery was determined to be the point at which that system had returned to within 99% of its original (pre-disrupted) performance. For our simulation, we were primarily interested in the impact of actual recovery time (RT) in water supply network on the simulated propagated recovery time (SPRT) in businesses network, which would then represent the difference in recovery times due to synchronization between federates in the SoS, as defined by time granularity – our key experimental parameter. In this way we could assess the accuracy of the SoS simulation and its dependence on time granularity.

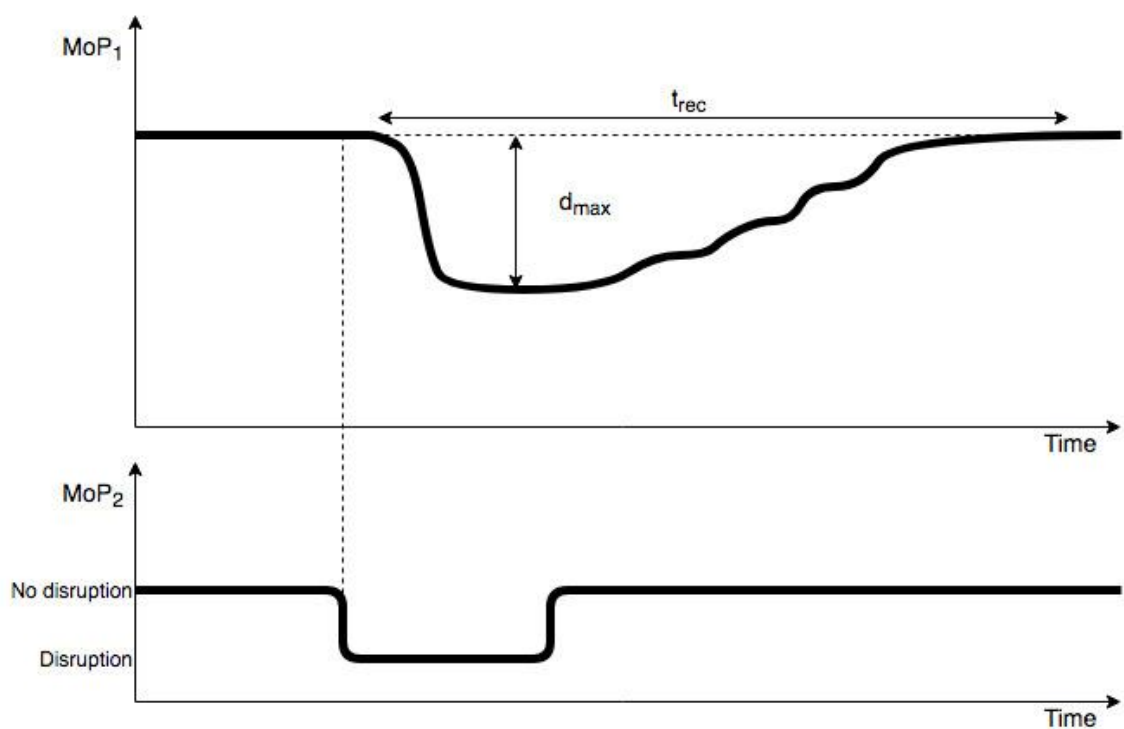

**Figure 5.** Recovery time ($t_{rec}$) and maximum disruption size ($d_{max}$) definitions. Disruption was introduced into water supply network (MoP2) and propagated to business network (MoP1).

Another dependent variable, maximum simulated propagated disruption size ($d_{max}$) (SPDS), was defined and measured as the lowest point along the performance curve after a disruption was propagated to that network. Those results indicated the magnitude of such an impact by one system on another, e.g., when a disruption to the water supply interfered with operations by the business network. This approach served as an alternative metric for assessing the accuracy of a simulation based on time granularity.

Our final evaluation involved comparing speeds (in seconds per timestep) among different time granularity configurations so that we could determine how the former changes in conjunction with the latter. This was an important factor because of the trade-off found between speed and accuracy in simulations. Successful design of a framework requires selecting the most appropriate time granularity based on desired speed and some basic knowledge about the networks being simulated. All of these were goals of our study here.

Finally, we devised a model of likelihood of visibility of disruption based on recovery time to time granularity ratio. Such model can help in selection of the optimal time granularity based on expected recovery time. Similarly, we developed a model that predicts estimated error for recovery time based on the relationship between simulated propagated recovery time and actual recovery time. This model can further aid with selection of an adequate time granularity. Moreover, such model could be used to estimate the actual recovery time of the original system based on the recorded recovery time in the propagated system.

*3.3. Model implementation*

Planners use distributed modelling frameworks to implement the SoS approach for businesses and infrastructure systems. This involves numerous individual, autonomous systems connected with each other through inputs and outputs to other systems. One such framework is HLA [38-40], a tool originated in military applications to simulate battlefield actions, as well as various systems pertaining to simulated battle situations and training. Since then, HLA has been employed in various other applications, including the modelling of civil infrastructure [25] [41-42]..

A particular implementation of HLA features three components: *interface specification*, *object model template (OMT)*, and *rules*. *Interface specification* defines where and how constituent systems ('federates') communicate with RTI, a method used to join all constituent distributed systems. This component serves as the inter-federate communication and synchronization unit of the HLA. Second, *OMT* describes what information is exchanged between constituent federates and what updates about the federation must be communicated to those federates. Finally, *rules* specify what federates have to obey to subscribe to the overall HLA SoS simulation ('federation'). When modelling infrastructure systems, the federates within an HLA simulation can include infrastructure systems, disruption generators, observers and visualization tools, patterns of user services, and businesses. All of these federates introduce dynamic changes to the systems and allow designers to observe their effects on the overall SoS.

Although HLA is perfectly suited to describing and modelling the manner in which constituent systems exchange information and synchronize with each other, it does not indicate the ideal time granularity at which such synchronization and exchange of information should take place. In this context, time granularity means the frequency of exchange of information among federates (i.e., constituent systems), disruption generators, and other components. As such, granularity reflects the frequency of inter-system synchronization. Although specification of HLA provides some mechanisms to perform time management [43-45], it does not identify the best time granularity for that exchange. In fact, the most adequate time granularity varies among types of simulations, where even different disruption events might apply to the same simulation of infrastructures.

Our simulation was developed in C++ v11 [46] and Python. The HLA framework was applied from Portico 2.0.2 HLA [47] implementation, with Portico's HLA being used to define the interfaces between federates, and to manage time in the simulation. Data at given time granularities were synchronized through HLA RTI, as adapted from Portico's implementation, and graph operations were performed with the use of the *igraph* library for C++ and Python, version 0.7.1 [48]. Disruptions were generated and introduced to the system through a disruption generator developed in C++ v11. All infrastructure system networks were developed in Python 3.5 [49], under Anaconda 2.4.0 distribution [50]. We used the following libraries to create those networks: *igraph*, for graph generation, representation, and operations; and *NumPy* version 1.10.1 [51], for linear algebra and numerical operations. The observer was designed with a webpage interface developed in JavaScript, HTML, and Python, using the *CanvasJS* library [52]. This observer enabled us to collect data about the simulations, to visualize their progress, and to view the performance of the system in real time. The simulation was developed, evaluated, and run on a Mac OS High Sierra 10.13.6 operating system.

This implementation shadowed a specific scheme. First, the infrastructure systems were developed based on the definitions established for their interconnections, number of nodes, and working mechanisms, i.e., inputs and outputs. Within each network, the nodes depended on preceding nodes and on their corresponding nodes from other federates. Once individual networks and their mechanisms were defined, the interfaces between federates were devised. This was followed by the design of HLA interfaces which considered what information and how often needs to be exchanged between federates. Finally, the overall HLA simulation was created by combining the individual constituents together to include all components of the infrastructure systems i.e. the three networks of interest.

Before arriving at our final experimental design, we evaluated the systems for different individual networks, each of which was tested to assess its representation of a real-life system. Our preliminary investigation showed that the networks and HLA SoS simulation performed well individually and as a whole, adequately representing individual networks and propagating and communicating disruptions between them as required. The networks were compared with simulations of individual infrastructures as devised by domain experts. This allowed us to assess the adequacy of the networks. A small-scale testing was developed to ensure that these networks respond well to disruptions being introduced. HLA SoS simulation as a whole was similarly tested with reference to expectations described by domain experts. Small-scale easy to estimate disruptions were introduced and response of the system was recorded. This followed the expectations of the experts.

Hence, the adequacy of the model was established. Furthermore, interdependencies introduced between federates in the SoS were evaluated by experts of power systems, water supply networks, and economists to ensure that these interdependencies corresponded well with the real-world and that duality of the simulation model was maintained.

In this section, we presented the experimental setup, layout and implementation of the model used in the simulation experiment. In the following section, we present the results of the experiments described in this section. Moreover, we provide an analysis of this results and derive statistical models based on our simulation results, which in turn can help us in making a recommendation to the optimal time granularity for the SoS simulation described in this section.

## 4. Results and analysis of the simulation experiment

In the previous section, we described the simulation experiment. In this section, we present and analyze the results of the simulation experiment performed in accordance with descriptions from the previous sections. Consequently, we make a recommendation for to the optimal time granularity for an SoS simulation of infrastructure systems and businesses.

### *4.1. Disruption size*

In this subsection, we evaluate the impact of various factors described in **Table 2** as well as time granularity, and interactions between these factors on the simulated, propagated disruption size (SPDS) performance metric described in Section 3.2. The experimental layout is shown in Section 3.1 and the implementation of the model for the simulation experiment executed here is explained in Section 3.3.

Simulated, propagated disruption size (SPDS) was measured under different experimental configurations to understand which factors had the largest impact on the simulated disruption size (SPDS). An ANCOVA (Analysis of Covariance) was performed in R to determine the strength of the effect of experimental factors on SPDS. **Figure 6** presents only those factors and their interactions that had a significant impact on the share in variation of SPDS. Adequate time granularity is of immense importance in system-of-systems models. The most important factor proved to be time granularity (TG), followed by actual recovery time (RT) and then actual disruption size (DS). We found this interesting for several reasons. First, the influence of propagated disruption was decided primarily by TG. Second, and more importantly, the RT was responsible for a greater share of the variation in SPDS than the DS. This finding demonstrated that the RT in an SoS federate had a greater effect on the SPDS in other federates than did the DS in the original federate itself especially for the case of three networks considered in this study. Although interactions among experimental factors also influenced the variations in SPDS, they had much less impact than did individual factors. Share of residuals in variation were also lower than the combined share of other factors. Overall, the experimental factors explained 71% of the variation of SPDS. That high percentage indicated that the variation in size could be well-explained by the experimental factors.

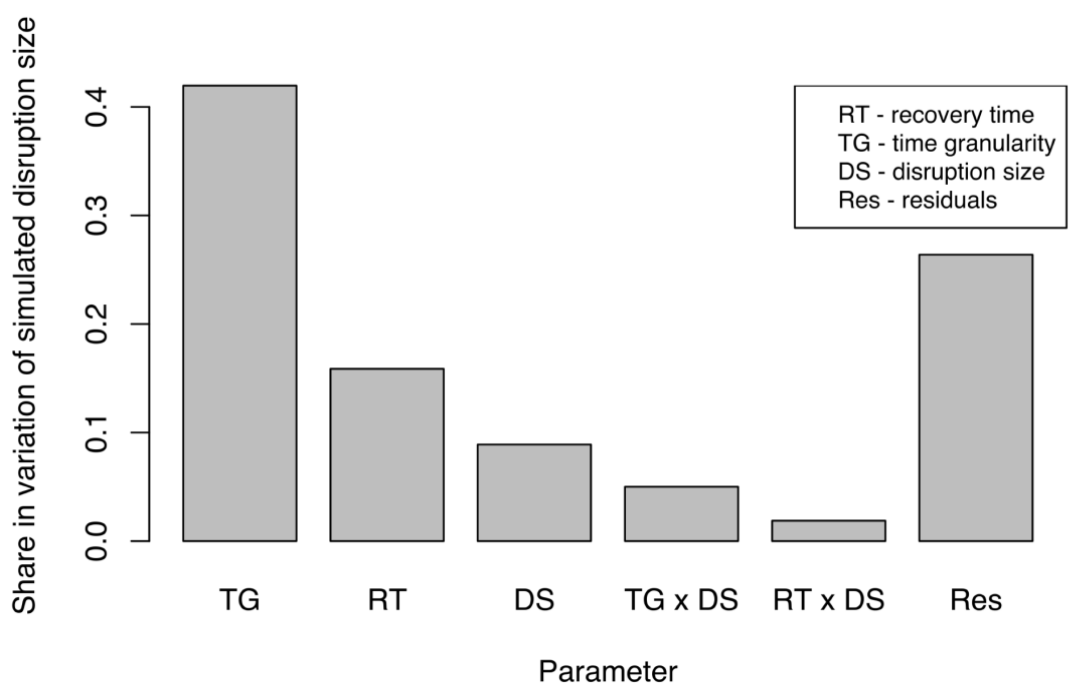

**Figure 6.** Share in overall variation for simulated disruption size (SPDS) due to experimental factors (parameters).

The results above suggested that correctly adjusting the ratio of time granularity to the expected recovery time in a simulation would be of immense importance when simulating the disruption size in an SoS evaluation. Therefore, careful selection of the ratio of time granularity to recovery time would have to be based on the data available to us if we were to determine the optimal time granularity for a simulation and yield the most accurate simulation results. These results signify the need for an adequate time granularity selection. Adjusting the time granularity can vary the outcomes of the simulation. Hence, when designing a simulation experiment time granularity needs to be considered. In particular, we need to make sure that the results with the time granularity selected are representing the real world accurately. As our further analysis below shows, the finer time granularity would yield more accurate results, and the coarser time granularity loses accuracy of the simulation.

*4.2. Recovery time*

In this subsection, we evaluate the impact of various factors described in **Table 2** as well as time granularity, and interactions between these factors on the simulated, propagated recovery time (SPRT) performance metric described in Section 3.2. The experimental layout is shown in Section 3.1 and the implementation of the model for the simulation experiment executed here is explained in Section 3.3.

Similar to our assessment of disruption size, we analyzed the impact of experimental factors on the variation in simulated, propagated recovery time (SPRT) of the business network. Simulated recovery time (SPRT) under various configurations was measured to understand how SPRT was affected by these factors. As before, we performed ANCOVA in R, using the data obtained when measuring the SPRT. **Figure 7** presents only the factors and their interactions that had significant impacts on the share in variation for SPRT. Impact was almost equally shared between time granularity (TG) and actual recovery time (RT). This indicated that both factors would require careful adjustments if simulations were to represent actual disruption events adequately. The most influential were time granularity (TG) of the simulation and the actual recovery time (RT) for the water supply network. Furthermore, TG had a slightly larger effect on SPRT. Again, this result underlined the importance of adequate TG and its critical influence on the accuracy and outcome of the simulation. The interaction between these two factors also had a significant but smaller share in the variation of SPRT. We noted with interest that the size of the disruption (DS) to the water supply had no significant impact on the SPRT. Similar to our results from examination of disruption sizes, the residual share in variation of SPRT was approximately 30%, which indicated that 70% of the variation (a reasonably high percentage) could be explained by experimental factors.

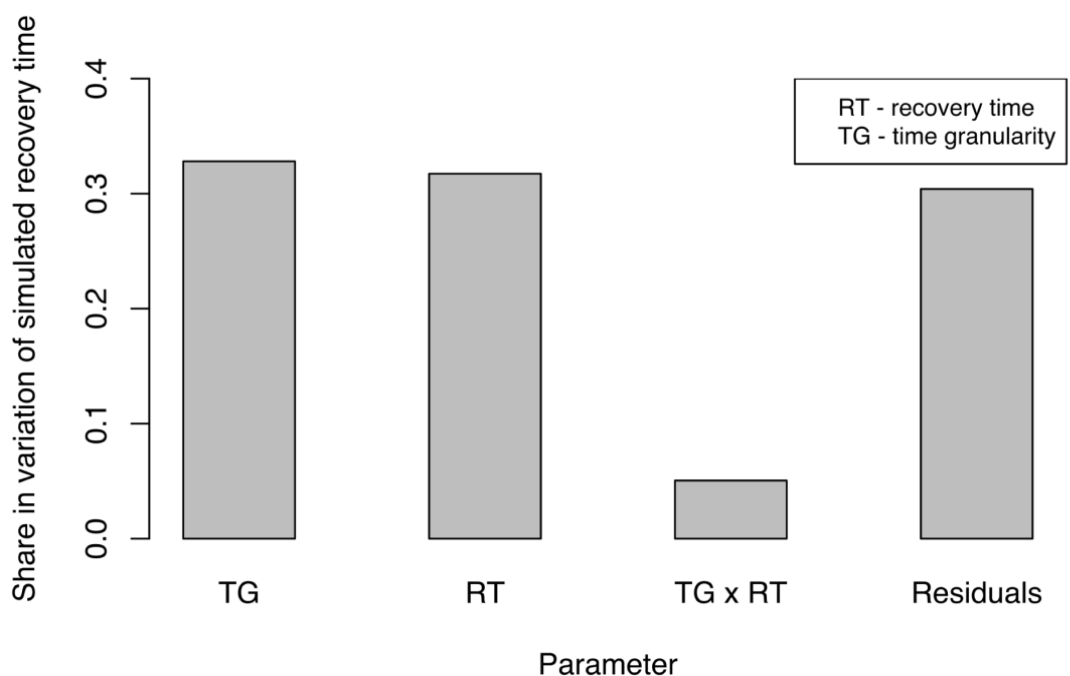

**Figure 7.** Variations in simulated recovery time (SPRT) due to experimental factors (parameters).

All of the findings presented above strongly suggested that time granularity (TG) was of great importance in an SoS simulation. Making proper adjustments to time granularity in reference to recovery time would be crucial if we were to obtain reliable simulation results. Furthermore, depending upon TG, simulation results could change significantly because the impact of RT and TG on the share of variation in SPRT was similar. In both cases of SPRT and SPDS, TG accounted for a larger share of the variation than did either the RT or the actual DS. Again, here we strongly see the importance of time granularity on obtaining adequate results of the simulation. The results obtained here signify that varying time granularity can result in outcomes, which do not represent real-world accurately, and in loss of propagation effects of disruptions. In particular, it is important to select time granularity which is fine enough to ensure that propagation of disruptions is noted in the simulation. The results obtained and described in this section support this view.

*4.3. Model of visibility of disruption*

One reason that time granularity and actual recovery time had such a great impact on the simulation outcome was that disruptions below a certain recovery time did not get propagated through the SoS simulation to other federates. If the time granularity was large enough, then when such a disruption occurred, the individual system recovered from the disruption before that propagation took place. This implied that time granularity prevented such an occurrence. As such, this phenomenon contributed to the great impact of time granularity on variation of the metrics.

Hence, in this subsection, we derive the optimal time granularity for the simulation of infrastructure systems and businesses described in Sections 2 and 3. This is done by running a logistic regression model on the experimental data. Namely, the ratio of RT to TG (described in **Table 2**) was used as the independent variable, and the dependent variable was measured considered as not-visible for SPDS lower than 5%, while visible if SPDS was greater than 5%. Such logistic regression model allows us to postulate the optimal time granularity for the simulation.

Based on these results, we then derived a model of likelihood for disruption visibility that would allow us to select time granularity of the simulation according to the (expected) recovery time of disruptions, making their propagation visible in the system. This model was used to select the maximum time granularity so that one could observe and detect an SoS disruption. Here, visibility of a disruption was understood to be a drop in performance to below 95% of the original level in the network to which disruption should propagate.

This logistical model (**Figure 8**) was developed based on our simulated data, using R and the *glm.fit* function [53]. It depicted the likelihood that a disruption would be visible in a business network, according to the ratio of actual recovery time of the water supply network to time granularity. The primary limitation associated with this model would be the networks to which it could be applied. Here, we used it to select a time granularity for a simulation based on the most important parameters: time granularity (TG) and actual recovery time (RT). Those parameters were combined into a ratio so that we could adjust for our desired likelihood of visibility. In most practical applications we can estimate the minimum actual recovery time. Then, by using the model from **Figure 8**, we can derive the time granularity. Our model suggested that the ratio of actual recovery time to time granularity should be at least 0.88. In doing so, for the disruption to propagate to other federates the time granularity should be less than 1.13 of the estimated actual recovery time.

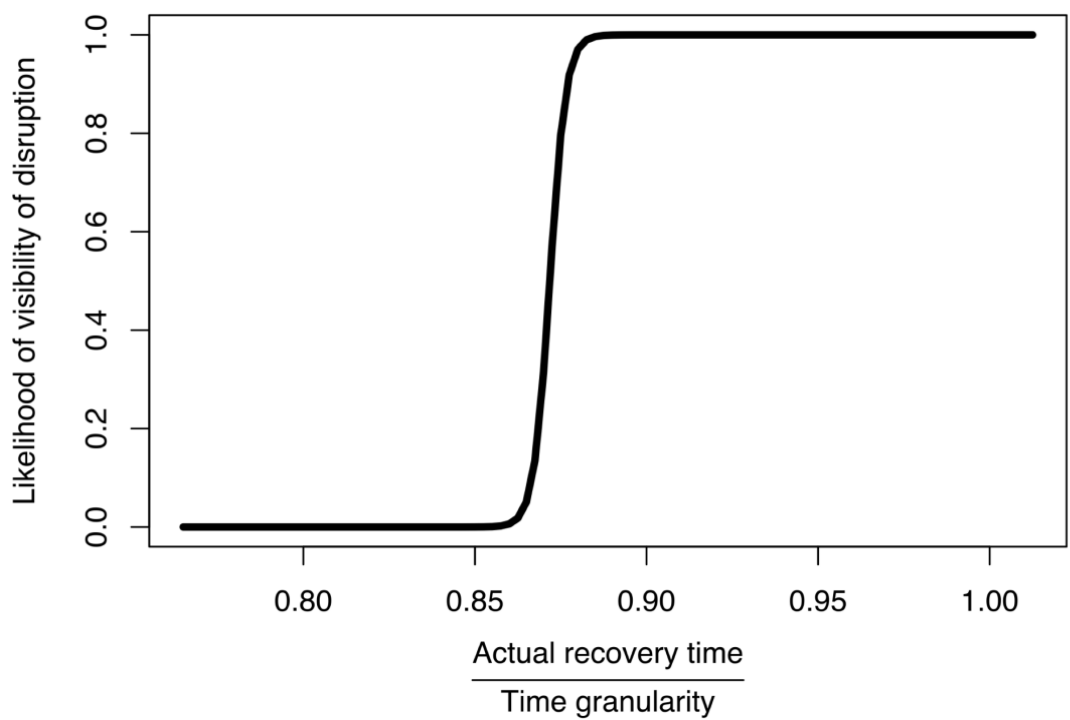

**Figure 8.** Model of likelihood for disruption visibility based on ratio of actual recovery time to time granularity (RT/TG).

Because visibility of a disruption is a key parameter when simulating an SoS infrastructure network, we wished to examine whether a disruption originating in one network could propagate to another network, for how long, and with what impact. This would enable us to understand how businesses respond to a certain disruption in the water supply network. However, if an excessively large time granularity prevented such a propagation, then the simulation would be useless.

Consequently, in a real-life scenario, we would not be able to determine whether the disruption would propagate. Hence, selection of appropriate time granularity for an SoS simulation is vital.

Our finding here could apply to multitude of other infrastructure networks, which also exhibit similar properties. These networks also require synchronization frequencies that are below or around expected recovery times. This is because individual networks otherwise might recover before the disruption is propagated. The general rule of maintaining recovery time above time granularity would hold. However, to obtain a detailed analysis of a different set of infrastructure networks, the simulation experiment should be conducted in a similar fashion with these particular networks following our general framework presented earlier. Such approach would allow to obtain a more accurate ratio of expected recovery time to time granularity that could apply for such networks. Similarly, uncertainty of the parameters could be estimated better with approaches such as bootstrapping and rerunning the simulation multiple times in order to obtain deviation from the numbers that we had obtained in the experiment. Stochasticity within each individual network and the overall system could be captured through such approach. However, this would require significant simulation time and might not be practical given the goals of such simulation. A representative sample of results used for calculation of the metric is sufficient in order to obtain a general rule for designing SoS of infrastructure systems and businesses.

Finally, the result here is of great significance, and such logistic model should be obtained when designing large simulation experiments for infrastructure systems. The time granularity needs to be fine enough to capture cascading effects of disruptions. Such model allows us to define the time granularity, which would be fine enough to ensure propagation of disruptions. This model is derived based on the previous two subsections (Section 4.1 and 4.2), which support the need for finding an adequate time granularity for a system-of-systems simulation. To respond to the need for obtaining an accurate simulation results, which represent real-world accurately, we have derived and presented the logistic model described in this section.

*4.4. Simulation time vs. time granularity trade-off*

In this subsection, we evaluate the impact of time granularity, as described in **Table 2** on the speed of execution of the simulation. This is important to note how speed is affected by time granularity to note that performance metrics described in Section 3.2 have a trade-off with the simulation speed. Both high speed and good visibility of disruption as described in the previous subsection being the desired outcome. Once, we know the maximum time granularity, we can use simulation time trade-off curve to further tweak and narrow down the time granularity of the simulation.

Because the SoS approach often entails performing one simulation immediately after another, we want to enable performing as many simulations as possible within a limited period of time to test different scenarios. Our goal was to achieve the most rapid simulation that was also the most accurate. Since simulation speed was affected by time granularity, at any given level of accuracy, a trade-off existed between time granularity and simulation time.

In our experiment, simulation time was expressed in seconds per timestep. As time granularity increased, the simulation time decreased because reduced granularities involved more synchronization between federates. Such synchronizations are time-consuming and computationally expensive, thereby lengthening the simulations. As shown in **Figure 9**, simulation time was roughly inversely proportional to time granularity. Nevertheless, we also noted that the greater the time granularity, the lower the accuracy of the simulation and the higher the likelihood that a disruption would not propagate to other networks – in our case, the businesses network. This produced a trade-off between simulation speed and accuracy that was controlled by the time granularity parameter. Consequently, when attempting to choose as high a time granularity as possible in order to decrease execution time, we still have to establish restrictions on maximum time granularity, so that disruption can effectively propagate between systems.

There is a clear traded-off between disruption time and time granularity of disruption. This can be seen based on results in this section. Hence, to estimate the time of running a simulation, we can use this model coupled with the minimum time granularity required obtained from the logistic model in Section 4.3. Furthermore, based on a given simulation budget, we can try use the below model to obtained the relevant time granularity and the experimental layout. Hence, it is important to perform such simulation time analysis and compare it with the time granularity requirements such as in previous sections.

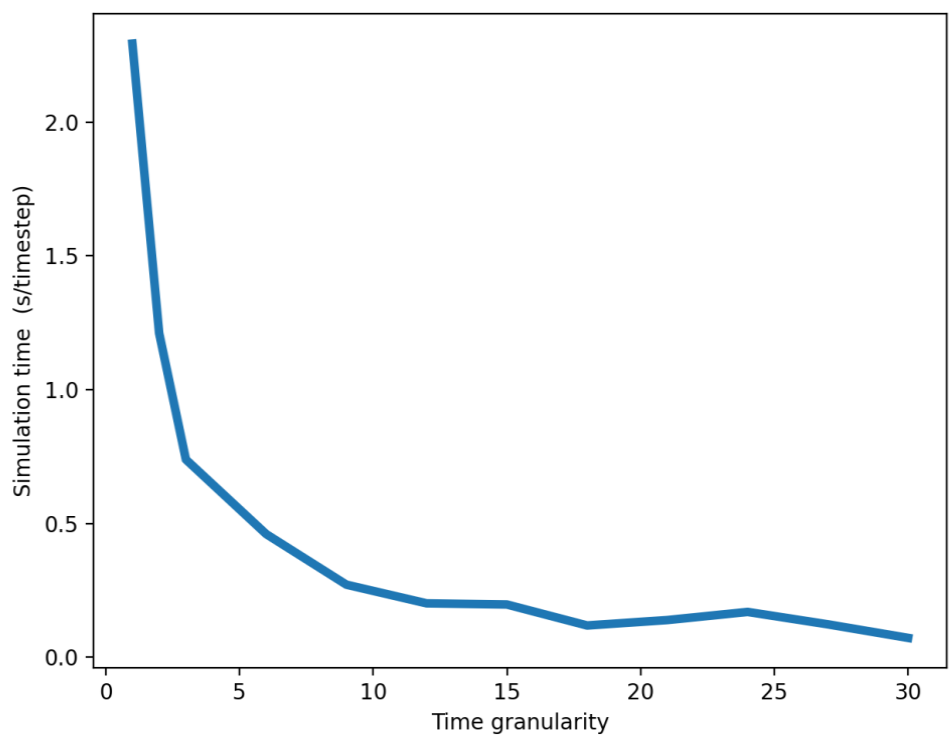

**Figure 9.** Inverse relationship that leads to trade-off between simulation time and accuracy, controlled by time granularity.

*4.5. Ratio of recovery time to time granularity*

In our SoS simulation, selecting the appropriate time granularity was critical. However, we detected a trade-off between time granularity and the speed of the simulation. The principal factors affecting our results were time granularity (TG) and recovery time (RT), both of which influenced the size of the simulated disruption to the business network (SPDS) more than did the size of the disruption to the water supply network (DS). Therefore, we concluded that the ratio of the recovery time for the original disruption to the water supply (i.e., actual recovery time (RT)) to the time granularity of the simulation (TG) had an even greater impact on the variation among our experimental metrics (both described in **Table 2**) on the simulation described in Section 3.

**Figure 10**, presents the results of our ANCOVA analysis. The ratio of actual recovery time of water supply network to time granularity (RT/TG) accounted for the greatest share, by far (~86%), in variation of the ratio of simulated recovery time of the business network to time granularity (SPRT/TG). This meant that the ratio RT/TG could be used to explain, with high accuracy, the ratio SPRT/TG. Therefore, SPRT/TG could be modelled based on RT/TG.

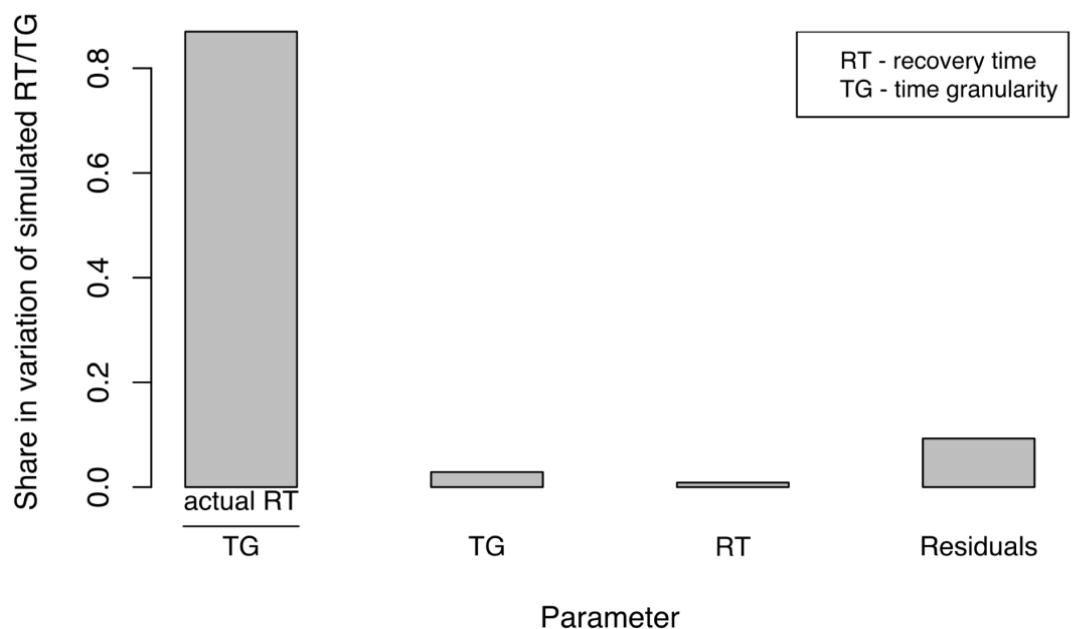

**Figure 10.** Variations in ratios of simulated recovery time by business network (SPRT) to time granularity due to experimental factors.

Consequently, we could use this model to estimate the desired accuracy of the simulation, this is the difference between SPRT and RT. Then, we can adjust TG accordingly to achieve this desired accuracy of the model.

One limitation associated with this model is its dependence on the type of simulation that is being run. Depending on the chosen federates, the actual difference in RT and SPRT recovery times might differ between actual conditions and those that are simulated or propagated. Therefore, the type of simulation, and its internal mechanics, as well as the type and number of interdependencies, have a great impact on this model's accuracy and validity. Likewise, Dubaniowski and Heinimann [6], have demonstrated that such a model that translates the ratio from actual (RT/TG) to simulated (SPRT/TG) values might have several regimes depending on whether the ratio is below or above 2. We found it also interesting that such a relationship did not emerge when using the disruption size metric. This further indicated that the selection of an appropriate time granularity is especially sensitive to the recovery time for the major event that is expected in a given simulation. Therefore, the minimum recovery time that is anticipated for major events should be carefully estimated when designing an SoS simulation.

Our proposed generalized linear model (**Figure 11**) was built to consider only the most significant factor, i.e., the ratio of recovery time to time granularity (RT/TG). It can be used to select the best time granularity depending on how much tolerance one has for error. Based on the maximum acceptable error, we can choose the highest time granularity that satisfies this to ensure the fastest simulation speed. Similarly, and perhaps more importantly, if we know the time granularity of a simulation, then the model can estimate the actual size of an event and its recovery time based on the simulated recovery time. This can help in obtaining the relevant data of a simulation that might have too coarse time granularity, which could be very beneficial, when we have limited simulation budget, but need to obtain as relevant results as possible. Using the linear model described in this section can be of essence to retrieving representative information from a simulation experiment with a given time granularity.

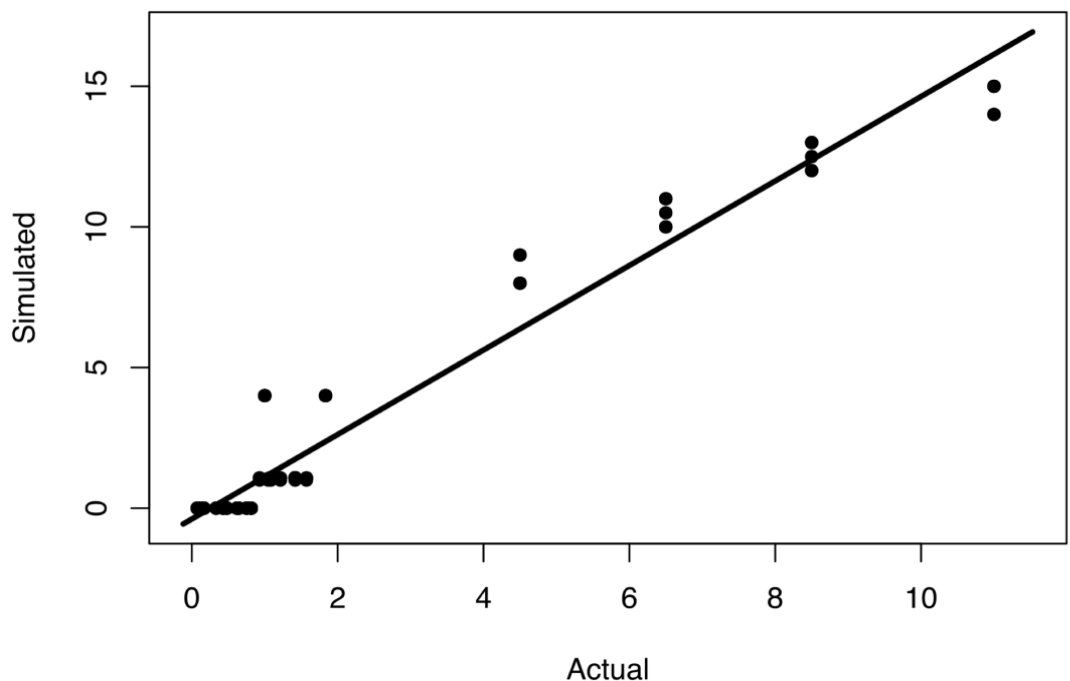

**Figure 11.** Model of relationship between simulated (SPRT) and actual (RT) recovery times as fraction of time granularity (TG).

We must reiterate our conclusion that this proposed model can vary in its applicability and accuracy, depending upon the networks that are simulated. Although we utilized networks here that resembled infrastructure systems such as water supply, power grid, and businesses, this model might not be fully applicable and yield different results in other situations. Nevertheless, the outcome from the study described here is valuable for deriving the actual size of a real event based on the simulated event size, and it can aid in choosing the appropriate time granularity, especially in SoS HLA simulations. However, a simulation following our general framework can be performed in order to find out more details about desired time granularities for different networks in an SoS simulation. Finally, it is possible that such a model could have several regimes depending upon the range of values of the ratio of actual recovery time to time granularity (RT/TG).

*4.6. Summary of key results*

Our research yielded the following major results:
- Time granularity had the greatest influence on both simulated recovery time and size of disruptions in systems to which those disruptions had propagated.
- Recovery time had a larger impact than disruption size on both recovery time and disruption size in systems to which disruptions had propagated.
- Experimental factors explained 70% of the variation in experimental performance metrics.
- From the model, we determined that the minimum ratio of actual recovery time to time granularity at which the propagation of disruptions was visible was 0.88.
- Simulation speed was inversely proportional to time granularity, and the best speeds were achieved at higher granularities.
- The ratio of actual recovery time to time granularity had the greatest effect on the ratio of simulated, propagated recovery time to time granularity. Hence, it was crucial to our simulation that we achieve an adequate ratio of actual recovery time to time granularity. The share of variation in simulated recovery time to time granularity ratio due to the ratio of actual recovery time to time granularity was 86%.

- We developed a general linear model to estimate the actual recovery time based on simulated, propagated recovery time.

## 5. Conclusions

Our simulation experiments were specifically designed to investigate how time granularity, recovery time, and size of a disruption to the water supply might propagate and affect the outcome of a business network under simulated conditions. After developing an HLA system-of-systems simulation that incorporated two infrastructure systems (water supply and power grid), a business network, and a disruption generator, we ran full-factorial simulation experiments to analyze various impacts on the propagation of disruptions. We then developed two models to assist in selecting an adequate time granularity based on expected recovery time and desired accuracy of the simulation.

The results from the simulation experiment demonstrated that time granularity, recovery time from disruptions, and disruption size all had a significant impact on the outcome of our simulation. Assessments of performance by the business network indicated that the simulated propagated recovery time (SPRT) and disruption size (SPDS) were affected by all experimental factors, but especially by time granularity and recovery time. The ratio of actual recovery time to time granularity had the greatest effect on the ratio of simulated, propagated recovery time to time granularity. Our study determined that the minimum ratio of actual recovery time to time granularity at which the propagation of disruptions was visible for infrastructure system networks and businesses was 0.88 Finally, we developed a general linear model to estimate the actual recovery time based on simulated, propagated recovery time.

Another important contribution of this study, besides the optimal ratio of time granularity to expected recovery time in an SoS simulation, is the development and proof-of-concept of the framework for deriving the optimal value of time granularity. This is a key contribution of this study to various fields of research that might benefit from distributed simulations of propagating disruptions including social economics, finance, sociology or biology to name a few.

Our study about the effect of time granularity on propagation of disruptions in a SoS simulation of infrastructure systems and businesses networks is novel. Our simulation experiment addressed the issue of selecting an appropriate time granularity when modelling disruptions in distributed modelling SoS environments such as the Portico HLA. This has closed a major gap in current research stream. Obtaining the suitable time granularity for such simulations will bring results of these SoS simulations closer to true values. Consequently, as a result of this study, practitioners, as well as decision-makers, of risk and disaster management will benefit from better simulations and decision support tools, which in turn would contribute to better responses to disruptions. In particular, the gap addressed in this study is the accurate selection of the time granularity parameter in an SoS simulation. Closing this gap will lead to more accurate simulations, and thus better decisions by the users of these simulations.

Our research findings have several further implications. Scientists can use the findings of this study to develop better models of infrastructure systems and business networks. The closing of the gap regarding selection of the appropriate time granularity will enable them to achieve better simulation representations at lower simulation costs. Scientists and researchers can select time granularity that yields the most optimal results of SoS simulations in terms of the most accurate disruption propagation representation. The experimental framework presented here can also be used to define time granularity of SoS simulations for applications other than infrastructure systems such as military systems, business systems or evolutionary dynamics of social economics. This would result in more accurate simulations in these fields thus decreasing the simulation time and budgets required, as well as yielding better results. For practitioners, policymakers, and scientists, better simulations can help them to achieve better estimates of the size and cost of actual disruptions. Thus, more accurate and faster simulations would result in better decisions regarding infrastructure development, contingency plans, and responses to ongoing disruptions.

Practitioners and scientists who regularly perform such simulations can benefit from identifying the maximum time granularity required if they are still to register the propagation of disruptions. This will aid in solving the trade-off between speed and accuracy of the simulation, which in turn would allow them to perform more simulations within the same timeframe, thereby leading to savings in both cost and time. Moreover, they can benefit from the framework to perform analysis of how individual factors affect propagation of disruptions between constituent systems in an SoS simulation of infrastructure systems. Addressing the gap of selecting the optimal time granularity, through faster and more accurate simulations will result in improved speed of simulations and so decreased time and budget spent on simulations. This is advantageous to scientists, practitioners, and decision-makers alike.

Our study is limited in its ability to apply these results to other networks and infrastructure systems. Although the test system was analyzed for various parameters of disruptions, the topologies and mechanics of the network were constant. In real-world scenarios, the nature of the networks, as well as their sizes, could differ. However, this can be remedied by applying our generalized framework to different situations and simulation in a small-scale, drawing conclusions from these to utilize when developing a large-scale simulation of the same systems. Similarly, the experimental space and metrics are restricted. Here, we focused on only a subset of possible factors that might influence a disruption. Another limitation was that the model of the ratio of simulated, propagated recovery time to time granularity based on the ratio of actual recovery time to time granularity may have several regimes that could be analyzed separately. Finally, our model was limited in capturing and representing stochasticity and uncertainty of results and parameters. To address these challenges, future work on this subject might include similar simulation experiments over a broader field of network topologies with different operating parameters. A real-life network could be used to investigate whether the results obtained here remain valid. Bootstrapping methods could be used to capture and represent uncertainty of results as a standard deviation of the resulting parameters. This would contribute to better representation and understanding of key results to make these more applicable. Finally, a wider range of disruption parameters could be studied, along with more experimental metrics, so that we could determine what other factors can affect the propagation of disruptions, and whether the models obtained through our study still apply.

**Author Contributions:** Conceptualization, M.D. and H.H.; methodology, M.D. and H.H.; software, M.D.; validation, M.D.; formal analysis, M.D.; investigation, M.D.; resources, M.D. and H.H.; data curation, M.D.; writing—original draft preparation, M.D.; writing—review and editing, M.D. and H.H.; visualization, M.D.; supervision, H.H.; project administration, M.D. and H.H.; funding acquisition, H.H. All authors have read and agreed to the published version of the manuscript

**Funding:** This research was funded by National Research Foundation (Singapore), grant number FI 370074011.

**Acknowledgments:** This work is an outcome of the Future Resilient Systems project at the Singapore-ETH Centre (SEC), which is funded by the National Research Foundation of Singapore (NRF) under its Campus for Research Excellence and Technological Enterprise (CREATE) programme. We would like to thank Priscilla Licht and Enago for help and support with language editing.

**Conflicts of Interest:** The authors declare no conflict of interest. The funders had no role in the design of the study; in the collection, analyses, or interpretation of data; in the writing of the manuscript, or in the decision to publish the results.

## References

1. Heinimann, H.R.; Hatfield, K. Infrastructure Resilience Assessment, Management and Governance – State and Perspectives. In *Resilience and Risk*; NATO Science for Peace and Security Series C: Environmental Security; Springer, Dordrecht, 2017; pp. 147–187 ISBN 978-94-024-1122-5.
2. Montanari, A.; Ratto, E.; Corsetti, E.; Morzenti, A. Embedding time granularity in logical specifications of real-time systems. In Proceedings of the Proceedings. EUROMICRO `91 Workshop on Real-Time Systems; 1991; pp. 88–97.

3. Wu, T.; Blackhurst, J.; O'grady, P. Methodology for supply chain disruption analysis. *International Journal of Production Research* **2007**, *45*, 1665–1682, doi:10.1080/00207540500362138.
4. Guo, S.; Hu, X.; Wang, X. On time granularity and event granularity in simulation service composition (WIP). In Proceedings of the Proceedings of the 2012 Symposium on Theory of Modeling and Simulation - DEVS Integrative M&S Symposium; Society for Computer Simulation International: Orlando, Florida, 2012; pp. 1–6.
5. Shi, J.; Wan, J.; Yan, H.; Suo, H. A survey of Cyber-Physical Systems. In Proceedings of the 2011 International Conference on Wireless Communications and Signal Processing (WCSP); 2011; pp. 1–6.
6. Dubaniowski, M.I.; Heinimann, H.R. Time Granularity in System-of-Systems Simulation of Infrastructure Networks. In Proceedings of the Complex Networks and Their Applications VII; Aiello, L.M., Cherifi, C., Cherifi, H., Lambiotte, R., Lió, P., Rocha, L.M., Eds.; Springer International Publishing, 2019; pp. 482–490.
7. Shi, Y.; Pan, M.; Peng, D. Replicator dynamics and evolutionary game of social tolerance: The role of neutral agents. *Economics Letters* **2017**, *159*, 10–14, doi:10.1016/j.econlet.2017.07.005.
8. Shi, Y.; Pan, M. Dynamics of Social Tolerance on Corruption: An Economic Interaction Perspective. *Journal for Economic Forecasting* **2018**, 135–141.
9. Shi, Y. Economic description of tolerance in a society with asymmetric social cost functions. *Economic Research-Ekonomska Istraživanja* **2019**, *32*, 2584–2593, doi:10.1080/1331677X.2019.1642784.
10. Tan, C.S.; Tan, P.S.; Lee, S.S.G.; Pham, M.T. An Inoperability Input-Output Model (IIM) for disruption propagation analysis. In Proceedings of the 2013 IEEE International Conference on Industrial Engineering and Engineering Management; 2013; pp. 186–190.
11. Han, J.; Shin, K. Evaluation mechanism for structural robustness of supply chain considering disruption propagation. *International Journal of Production Research* **2016**, *54*, 135–151, doi:10.1080/00207543.2015.1047977.
12. Scheibe, K.P.; Blackhurst, J. Supply chain disruption propagation: a systemic risk and normal accident theory perspective. *International Journal of Production Research* **2018**, *56*, 43–59, doi:10.1080/00207543.2017.1355123.
13. Eusgeld, I.; Nan, C. Creating a simulation environment for critical infrastructure interdependencies study. In Proceedings of the 2009 IEEE International Conference on Industrial Engineering and Engineering Management; 2009; pp. 2104–2108.
14. Rinaldi, S.M.; Peerenboom, J.P.; Kelly, T.K. Identifying, understanding, and analyzing critical infrastructure interdependencies. *IEEE Control Systems Magazine* **2001**, *21*, 11–25, doi:10.1109/37.969131.
15. Boer, C. *Distributed Simulation in Industry*; Erasmus University Rotterdam, 2005; ISBN 978-90-5892-093-5.
16. Guidotti, R.; Gardoni, P.; Rosenheim, N. Integration of physical infrastructure and social systems in communities' reliability and resilience analysis. *Reliability Engineering & System Safety* **2019**, *185*, 476–492, doi:10.1016/j.ress.2019.01.008.
17. Rinaldi, S.M. Modeling and simulating critical infrastructures and their interdependencies. In Proceedings of the Proceedings of the 37th Annual Hawaii International Conference on System Sciences, 2004.; Big Island, HI, USA, 2004.
18. Klein, U.; Schulze, T.; Straßburger, S. Traffic Simulation Based on the High Level Architecture. In Proceedings of the Proceedings of the 30th Conference on Winter Simulation; IEEE Computer Society Press: Los Alamitos, CA, USA, 1998; pp. 1095–1104.
19. Liu, X.; Liu, J.; Zhao, S.; Tang, L.C. Modeling and Simulation on a Resilient Water Supply System Under Disruptions. In *Proceedings of the Institute of Industrial Engineers Asian Conference 2013*; Springer, Singapore, 2013; pp. 1385–1393 ISBN 978-981-4451-97-0.
20. Kinney, R.; Crucitti, P.; Albert, R.; Latora, V. Modeling cascading failures in the North American power grid. *Eur. Phys. J. B* **2005**, *46*, 101–107, doi:10.1140/epjb/e2005-00237-9.
21. Moulsley, T.J.; Lo, P.; Haddon, J.; Vilar, E. The efficient acquisition and processing of propagation statistics. *Journal of the Institution of Electronic and Radio Engineers* **1985**, *55*, 97–103, doi:10.1049/jiere.1985.0034.
22. Morlet, J. Sampling Theory and Wave Propagation. In *Issues in Acoustic Signal — Image Processing and Recognition*; Chen, C.H., Ed.; Springer Berlin Heidelberg: Berlin, Heidelberg, 1983; pp. 233–261 ISBN 978-3-642-82004-5.
23. Hassab, J.C. *Underwater Signal and Data Processing*; CRC Press, 2018; ISBN 978-1-351-07744-6.
24. Marziale, N.; Nobilia, F.; Pellegrini, A.; Quaglia, F. Granular Time Warp Objects. In Proceedings of the Proceedings of the 2016 ACM SIGSIM Conference on Principles of Advanced Discrete Simulation; Association for Computing Machinery: New York, NY, USA, 2016; pp. 57–68.
25. Eusgeld, I.; Nan, C.; Dietz, S. "System-of-systems" approach for interdependent critical infrastructures. *Reliability Engineering & System Safety* **2011**, *96*, 679–686, doi:10.1016/j.ress.2010.12.010.

26. Dueñas-Osorio, L.; Vemuru, S.M. Cascading failures in complex infrastructure systems. *Structural Safety* **2009**, *31*, 157–167, doi:10.1016/j.strusafe.2008.06.007.
27. Dubaniowski, M.I.; Heinimann, H.R. A framework modeling flows of goods and services between businesses, households, and infrastructure systems. In Proceedings of the Resilience The 2nd International Workshop on Modelling of Physical, Economic and Social Systems for Resilience Assessment : 14-16 December 2017, Ispra; Publications Office of the European Union: Luxembourg, 2017; Vol. I, pp. 182–190.
28. Dubaniowski, M.I.; Heinimann, H.R. A framework for modeling interdependencies among households, businesses, and infrastructure systems; and their response to disruptions. *Reliability Engineering & System Safety* **2020**, *203*, 107063, doi:10.1016/j.ress.2020.107063.
29. Aydin, N.Y.; Duzgun, S.; Heinimann, H.R. Resilience evaluation for transportation networks accessibility under seismic risk.; Adelaide, Australia, 2017.
30. Chelst, K.R.; Barlach, Z. Multiple Unit Dispatches in Emergency Services: Models to Estimate System Performance. *Management Science* **1981**, *27*, 1390–1409, doi:10.1287/mnsc.27.12.1390.
31. Tang, J.; Khoja, L.; Heinimann, H.R. Characterisation of survivability resilience with dynamic stock interdependence in financial networks. *Applied Network Science* **2018**, *3*, 23, doi:10.1007/s41109-018-0086-z.
32. Gao, J.; Buldyrev, S.V.; Stanley, H.E.; Havlin, S. Networks formed from interdependent networks. *Nature Physics* **2012**, *8*, 40–48, doi:10.1038/nphys2180.
33. Eid, M. A general analytical solution for the occurrence probability of a sequence of ordered events following Poisson stochastic processes. *Reliability: Theory & Applications* **2011**, *6*.
34. Eid, M.; Serafin, D.; Barbarin, Y.; Kuligowska, E.; Soszyńska-Budny, J.; Kolowrocki, K. A resilience model based on Stochastic Poison Process.; 2015; pp. 21–27.
35. Olsson, A.M.J.; Sandberg, G.E. Latin Hypercube Sampling for Stochastic Finite Element Analysis. *Journal of Engineering Mechanics* **2002**, *128*, 121–125, doi:10.1061/(ASCE)0733-9399(2002)128:1(121).
36. Erdős, P.; Rényi, A. On Random Graphs I. *Publicationes Mathematicae (Debrecen)* **1959**, *6*.
37. Bollobás, B. Random Graphs. In *Modern Graph Theory*; Bollobás, B., Ed.; Graduate Texts in Mathematics; Springer New York: New York, NY, 1998; pp. 215–252 ISBN 978-1-4612-0619-4.
38. Topçu, O.; Oğuztüzün, H. High Level Architecture. In *Guide to Distributed Simulation with HLA*; Topçu, O., Oğuztüzün, H., Eds.; Simulation Foundations, Methods and Applications; Springer International Publishing: Cham, 2017; pp. 29–78 ISBN 978-3-319-61267-6.
39. Dahmann, J.S.; Fujimoto, R.M.; Weatherly, R.M. The Department of Defense High Level Architecture. In Proceedings of the Proceedings of the 29th conference on Winter simulation - WSC '97; ACM Press: Atlanta, Georgia, United States, 1997; pp. 142–149.
40. Dahmann, J.S.; Kuhl, F.; Weatherly, R. Standards for Simulation: As Simple As Possible But Not Simpler The High Level Architecture For Simulation. *SIMULATION* **1998**, *71*, 378–387, doi:10.1177/003754979807100603.
41. Ferenci, S.L.; Choi, M.; Evans, J.; Fujimoto, R.M.; Alspaugh, C.; Legaspi, A.K. Experiences integrating NETWARS with the naval simulation system using the high level architecture. In Proceedings of the IEEE MILCOM 2004. Military Communications Conference, 2004.; 2004; Vol. 3, pp. 1395-1401 Vol. 3.
42. Jain, A.; Robinson, D.; Dilkina, B.; Fujimoto, R. An approach to integrate inter-dependent simulations using HLA with applications to sustainable urban development. In Proceedings of the 2016 Winter Simulation Conference (WSC); IEEE: Washington, DC, USA, 2016; pp. 1218–1229.
43. Fujimoto, R.M. Time Management in The High Level Architecture. *SIMULATION* **1998**, *71*, 388–400, doi:10.1177/003754979807100604.
44. Fujimoto, R.M.; Weatherly, R.M. Time Management in the DoD High Level Architecture. In Proceedings of the Proceedings of the Tenth Workshop on Parallel and Distributed Simulation; IEEE Computer Society: Washington, DC, USA, 1996; pp. 60–67.
45. Jansen, R.E.J.; Huiskamp, W.; Boomgaardt, J.J. Real-time Scheduling of HLA Simulator Components. *TNO Fysisch en Elektronisch Laboratorium* **2004**.
46. Josuttis, N.M. *The C++ Standard Library: A Tutorial and Reference*; Addison-Wesley Professional, 2012; ISBN 978-0-321-62321-8.
47. Calytrix Technologies The Portico Project Available online: http://www.porticoproject.org/comingsoon/ (accessed on Feb 11, 2019).
48. igraph.org python-igraph 0.7.1 Available online: http://igraph.org/2014/02/04/igraph-0.7-python.html (accessed on Aug 7, 2018).

49. Python.org Python Release Python 3.5.0 Available online: https://www.python.org/downloads/release/python-350/ (accessed on Aug 7, 2018).
50. Anaconda Inc. Anaconda 2.4.0 for Python 3.5 Available online: https://docs.anaconda.com/anaconda/packages/old-pkg-lists/2.4.0/py35 (accessed on Aug 7, 2018).
51. SciPy.org NumPy Reference — NumPy v1.10 Manual Available online: https://docs.scipy.org/doc/numpy-1.10.1/reference/ (accessed on Aug 7, 2018).
52. Fenopix, Inc. CanvasJS 1.8.5 beta Available online: https://canvasjs.com/ (accessed on Aug 7, 2018).
53. Lewis, B. An R Interface to SciDB Available online: https://www.rdocumentation.org/packages/scidb/versions/1.2-0 (accessed on Feb 8, 2019).